\documentclass[11pt]{article}

\renewcommand{\d}{\partial}

\def\slashed#1{\slash\hskip-6pt#1}
\def\Nslash{\slash\hskip-8pt N}

\newcommand{\be}{\begin{equation}} 
\newcommand{\ee}{\end{equation}} 
\newcommand{\bea}{\begin{eqnarray}} 
\newcommand{\eea}{\end{eqnarray}} 
\newcommand{\bref}[1]{(\ref{#1})} 
\newcommand{\beasn}{\begin{sneqnarray}} 
\newcommand{\eeasn}{\end{sneqnarray}}

\newfont{\nice}{eufm10 scaled\magstep1}

\makeatletter 
 
\def\dif{{\rm d}} 
\def\deriv{\@ifnextchar[{\@deriv}{\@deriv[]}} 
   \def\@deriv[#1]#2#3{\mathchoice%
{{\dif^{#1}#2\over\dif{#3}^{#1}}}{{\dif^{#1}#2/\dif{#3}^{#1}}}%
{{\dif^{#1}#2\over\dif{#3}^{#1}}}{{\dif^{#1}#2/\dif{#3}^{#1}}}}

\def\secteqno{\@addtoreset{equation}{section}%
\def\theequation{\thesection.\arabic{equation}}} 
\def\endsecteqno{\def\theequation{\@ifundefined{chapter}%
{\arabic{equation}}{\thechapter.\arabic{equation}}}} 
\newcounter{subequation} 
\def\thesubequation{\alph{subequation}} 
\def\sneqnarray{\stepcounter{equation}\let\@currentlabel=\theequation 
\setcounter{subequation}{1} 
\def\@eqnnum{{\rm (\theequation.\thesubequation)}} 
\global\@eqcnt\z@\tabskip\@centering\let\\=\@eqncr\let\@@eqncr=\@@sneqncr 
$$\halign to \displaywidth\bgroup\@eqnsel\hskip\@centering 
 $\displaystyle\tabskip\z@{##}$&\global\@eqcnt\@ne 
 \hskip 2\arraycolsep \hfil${##}$\hfil 
 &\global\@eqcnt\tw@ \hskip 2\arraycolsep $\displaystyle\tabskip\z@{##}$\hfil 
  \tabskip\@centering&\llap{##}\tabskip\z@\cr} 
\def\endsneqnarray{\@@sneqncr\egroup $$\global\@ignoretrue} 
\def\@@sneqncr{\let\@tempa\relax 
   \ifcase\@eqcnt \def\@tempa{& & &}\or \def\@tempa{& &} 
   \else \def\@tempa{&}\fi 
     \@tempa \if@eqnsw\@eqnnum\stepcounter{subequation}\fi 
     \global\@eqnswtrue\global\@eqcnt\z@\cr} 
\def\nobiblabels{\def\@lbibitem[##1]##2{\@bibitem{##2}}} 
\makeatother 
 
\newcommand{\NPB}[3]{{\sl Nucl. Phys.} {\bf B#1} (#2) #3}

\newcommand{\PRevD}[3]{{\sl Phys. Rev.} {\bf D#1} (#2) #3}
\newcommand{\PLB}[3]{{\sl Phys. Lett.} {\bf #1B} (#2) #3}
\newcommand{\MPLA}[3]{{\sl Mod. Phys. Lett.} {\bf A#1} (#2) #3}

\newcommand{\PRSA}[3]{{\sl Proc. Roy. Soc.} {\bf A#1} (#2) #3}

\title{\LARGE\sffamily 
       Covariant Derivation of Effective Actions\\
       for SUSY Topological Defects
       \vspace{3mm}}

\author{\sc{Jordi Par{\'\i}s$^\sharp$ and Jaume Roca$^\flat$}
        \vspace{2mm}\\
        \it{Departament d'Estructura i Constituents
               de la Mat{\`e}ria}\\
        \it{Universitat de Barcelona, Diagonal 647}\\
        \it{E-08028 BARCELONA}
        \vspace{1mm}\\
        {\small $^{(\sharp)}$ \tt paris@ecm.ub.es}\\ 
        {\small $^{(\flat)}$ \tt roca@ecm.ub.es}
       }

\date{}

\begin{document}

\secteqno

\maketitle

\vskip -100mm
\vbox{\hfill\large UB-ECM-PF 98/05\null\par
      \hfill March 1998\null\par
      \hfill {\tt hep-th/9803037}}\null
\vskip  85mm

\begin{abstract}

We make a first step to extend to the supersymmetric arena the
effective action method, which is used to covariantly deduce the low
energy dynamics of topological defects directly from their parent
field theory.  By focussing on two-dimensional supersymmetric theories
we are able to derive the appropriate change of variables that singles
out the low energy degrees of freedom. These correspond to
super-worldline embeddings in superspace which are subject to a
geometrical constraint.  We obtain a supersymmetric and
$\kappa$--invariant low energy expansion, with the standard
superparticle action as the leading term, which can be used for the
determination of higher-order corrections.  Our formulation fits quite
naturally with the present geometrical description of
$\kappa$--symmetry in terms of the so-called geometrodynamical
constraints. It also provides a basis for the exploration of these
issues in higher-dimensional supersymmetric theories.

\end{abstract}

\newpage

\section{Introduction}

It is a well known fact that, at sufficiently low energies, the
dynamics of topological defects of bosonic relativistic field theories
is governed by Dirac-Nambu-Goto--type effective actions \cite{Dirac}.
Adding supersymmetry amounts to the replacement of such actions by
their supersymmetric extensions, the so-called super $p$-brane
actions, with the common feature of being invariant under a fermionic
gauge transformation, the (kappa) $\kappa$--symmetry \cite{PKT}.
Together with the standard reparametrization invariance, the
$\kappa$--symmetry allows to gauge away the redundant degrees of
freedom that are necessarily present for a manifestly Poincar{\'e}
invariant description of the effective dynamics of these objects.

Two different, although complementary, lines have been pursued in
order to identify the low energy dynamics of topological defects.  A
first approach --which rests on general principles of effective
actions and symmetries-- consists essentially in identifying the most
general action, with the lowest possible number of derivatives,
compatible with the required field content and symmetries of the low
energy regime of the theory.  In this way, a complete classification
of the possible super $p$--brane actions is presently at our disposal
\cite{Clas}.

In the second approach instead one attempts to deduce the low energy
effective action {\it directly} from the underlying field theory by
trying to single out the degrees of freedom which are relevant in this
regime and integrating the rest.  This will generally be a lengthier
and more intricate way to proceed because of the difficulty involved
in this splitting procedure. Yet one expects as a reward to be able to
understand better the origin of the low energy symmetries ---the
$\kappa$--symmetry in the case of supersymmetric theories--- and to
get hints from it on how to characterize these symmetries in a fully
geometrical way. Furthermore, this method will furnish a perturbative
expansion that can be used to get higher-order corrections to the
basic Dirac-Nambu-Goto--type actions.

The origin of this approach can be traced back to the covariant method
developed by F{\"o}rster \cite{F84} to obtain the effective dynamics of
the Nielsen-Olesen vortices \cite{NOV}. A key ingredient of this
method consists in coordinating spacetime by means of a set of
curvilinear coordinates adapted to the topological defect. These new
coordinates are composed of the worldsheet parameters, giving the
spacetime location of the topological defect, plus normal coordinates
parametrizing orthogonal displacements with respect to it. While for
bosonic topological defects the covariant derivation of the low energy
effective actions through this method is fairly well understood,
preliminary proposals in supersymmetric cases \cite{IvaKap1}, however,
have been able to produce only gauge-fixed versions of the
$\kappa$--invariant super $p$--brane actions. Clearly thus, the
supersymmetric extension of this procedure still lacks a proper
parametrization of the low energy degrees of freedom, {\it i.e.}~a
proper supersymmetry-invariant generalization of the above
defect-adapted coordination.

In fact, this central obstruction for a systematic derivation of
supersymmetric low energy effective actions is closely related to
another long-standing problem in this framework: the search for a
tensor calculus for the local $\kappa$--symmetry, known to be a
guiding ingredient in the construction of higher-order ``curvature''
corrections to the basic super $p$--brane actions.  Several attempts
have been made in order to develop such a calculus
\cite{STV,Gauntlett,IvaKap2,BSTPV}.  The common trend has been to
impose a so-called {\it geometrodynamical constraint\/} on the basic
superembedding describing the location of the topological defect in
superspace.  In this way $\kappa$--symmetry transformations arise as
remnants, from within the larger set of general superdiffeomorphisms,
that preserve the geometrodynamical constraint.  This approach has
definitely provided an important insight over the nature of
$\kappa$--symmetry. However, it is also clear that a full
understanding of the subject would involve explaining the origin of
this constraint and studying its relevance in the derivation of the
effective action in terms of the underlying field theory.

In this paper, we make a first step to answer the above questions by
addressing them in the framework of a generic supersymmetric scalar
field theory in two dimensions. We review in Sect.~2 the main features
of the covariant effective action method in the simpler bosonic
case. In Sect.~3 we introduce the supersymmetric theory and in Sect.~4
we extend the method to this case. A suitable generalization is
derived of the change of variables that we use to single out the low
energy modes. Consistency of this change, however, requires an extra
condition on the superembedding, implying that only a special kind of
superembeddings will be able to represent a super domain wall.  This
condition turns out to coincide with the normal projection of the
geometrodynamical constraint.  Imposing the tangent projection as well
simply amounts to a gauge-fixing reducing the field content and gauge
symmetry (superdiffeomorphisms) down to the standard one, {\it
i.e.}~reparametrizations and $\kappa$--symmetry. In the end, by
straightforward application of the derived change of variables, the
lowest-order effective action is directly obtained from the original
field theory. We conclude in Sect.~5 with a few remarks on the
extension of this work to higher-dimensional supersymmetric models.

\section{Effective actions for bosonic topological defects}

In this section we briefly describe the method used to obtain
effective actions describing the low energy regime of topological
defects in the simplest situation, {\it i.e.}~that of soliton-like
configurations in a $2d$ scalar field theory.  This will help us
illustrate the basic steps to follow later in the derivation of the
effective action in the supersymmetric case.

Consider a bosonic two--dimensional model described by a generic
action of the form
\begin{equation}
S[\varphi]=\int d^2x\left[\frac12(\partial_\mu\varphi)^2
-\frac12(U'(\varphi))^2\right].
\label{boseact}
\end{equation}
Such a model features soliton configurations whenever the positive
semi-definite potential given by $U'(\varphi)^2/2$ possesses a
non-trivial vacuum structure, {\it i.e.\ }when there are two or more
constant field configurations for which the potential assumes its
minimum zero value.  Typical examples are, for instance, the
$\varphi^4$ kink, where
$U'={\lambda\over2}\left(\varphi^2-{m^2\over\lambda^2}\right)$ gives
rise to two different vacua, $\varphi=\pm m/\lambda$, and the
sine-Gordon soliton, where $U'={2m^2\over\lambda} \sin\left({\lambda
\varphi\over 2 m}\right)$, with an infinite set of vacua \cite{Raj}.

Lowest energy solutions of this kind can be obtained by looking for
static configurations interpolating between two different vacua, say
$\varphi_{0+}$ and $\varphi_{0-}$, {\it i.e.} satisfying the boundary
conditions
\begin{equation}
\varphi\rightarrow \varphi_{0\pm}\quad\quad 
\mbox{\rm for}\quad\quad x\rightarrow\pm\infty.
\label{topbc}
\end{equation}
Then, it is clear that this sort of solutions will be stable against
decay into any of the vacua of the model because such transitions
would involve an infinite amount of energy.

For a static soliton the equation of motion derived from
\bref{boseact} reduces to
\begin{equation} 
-{d^2\varphi\over d x^2}+U'U''=0,\quad\quad
\Rightarrow\quad\quad\varphi'=\pm U'(\varphi),
\label{bosequ}
\end{equation}
together with boundary conditions of the form \bref{topbc}.  Due to
the time-independence of its solutions, $\varphi_{S,A}(x-a)$
(corresponding to soliton $(+)$ and antisoliton $(-)$, respectively),
it is said that they break only half of the translational symmetries
of the model. The space coordinate is the broken direction because of
the explicit $x$-dependence, while time remains as the unbroken
direction.

The specific boundary conditions obeyed by these soliton
configurations give them the common feature of dividing space in two
zones such that in each of them $\varphi_{S,A}(x-a)$ is basically in one
of the vacua. Both zones will meet in a narrow region around $x=a$
where the field rapidly evolves from one vacuum to the other and where
its energy is effectively confined ---a generically called domain
wall.  These stable energy lumps can thus be associated with
particle-like objects, the width of the transition region between
vacua being proportional to the inverse of the particle mass $M_S$.
In a low energy regime, characterized by $E<<M_S$, non-zero modes will
scarcely be excited because their typical scale is of the order
$M_S$. This implies that an effective description of the dynamics of
the system in this sector can naturally be achieved by focussing on
the dynamics of the zero modes alone, {\it i.e.} in terms of the
location of the soliton regarded as a point particle.  To study such
dynamics, one may consider small perturbations around the above
soliton solutions which will generally make the domain wall fluctuate
while approximately preserving the shape of the $\varphi$-field
configuration.

\medskip

Let us now review the process of derivation of the effective
action $S_{\rm eff}[x(t)]$, where $x^\mu(t)$ is the worldline of the
soliton center, and $t$ represents an arbitrary parametrization of
this worldline.  Essentially, one has to 
perform a splitting in the degrees of freedom contained in $\varphi$
between the zero modes, describing the location $x^\mu(t)$ of the 
soliton, and the other (massive) modes,
\begin{equation}
\varphi\quad\rightarrow\quad(x^\mu(t),\phi),
\label{split}
\end{equation}
and integrate then over the massive modes $\phi$.

A common way to do this splitting consists in first making a change of
variables from standard spacetime coordinates $(x^\mu)$ to a new pair
of coordinates $(t,\rho)$, such that both sets are related by
\begin{equation}
x^\mu=x^\mu(t)+\rho\, n^\mu(t),
\label{bosechange}
\end{equation}
where $x^\mu(t)$ is the worldline describing the location of the
soliton and $n^\mu(t)$ the unit normal to the unit tangent vector
$v^\mu(t)$.  The splitting \bref{split} is afterwards implemented by
defining $\phi(t,\rho)=\varphi(x^\mu)$ by means of this change of
variables.

In fact, the rationale behind the change \bref{bosechange} is
precisely to provide a covariant splitting between instantaneous
broken and unbroken symmetry directions.  With this idea in mind, the
form \bref{bosechange} of the change of variables can be directly
guessed but it can also be derived with the help of a simple method.
The procedure consists in, first, finding the transformation from the
$(x^\mu)$ spacetime coordinates to the coordinates $(z^\mu)$ of the
instantaneous co-moving frame at, say, $t=t_0$; and, second, writing
the instantaneous space direction in terms of the original $(x^\mu)$
coordinates.  Both frames $(x^\mu)$ and $(z^\mu)$ will be related by a
Poincar{\'e} transformation
\begin{equation}
x^\mu=\Lambda^\mu_{\;\nu}(t_0)z^\nu+a^\mu(t_0),
\label{poinctrans}
\end{equation}
where the values of $\Lambda^\mu_{\;\nu}(t_0)$ and $a^\mu(t_0)$ are
determined by imposing that $z(t)$ ---the expression of the worldline
$x^\mu(t)$ in the $(z^\mu)$ coordinates--- should have an expansion
around $t=t_0$ 
$$
z^\mu(t)=z^\mu(t_0)+\dot z^\mu(t_0)(t-t_0)+ O((t-t_0)^2),
$$
such that, at first order, the particle is at rest in the origin, {\it
i.e.}
$$
z^\mu(t_0)=0,\quad\quad
\dot z^\mu(t_0)\propto\delta^{\mu0}.
$$
It is a simple exercise to check that these conditions determine the
parameters of the Poincar{\'e} transformation \bref{poinctrans} to be
\begin{equation}
(\Lambda^\mu_{\;\nu}(t_0))=(v^\mu(t_0)~n^\mu(t_0)),\quad\quad
a^\mu(t_0)=x^\mu(t_0).
\label{poincparam}
\end{equation}
Now, locally around $t=t_0$, the instantaneous broken symmetry
direction is described by the pure space direction in the $(z^\mu)$
coordinates, {\it i.e.}~the spacetime points of the form
$z^\mu=\rho\,\delta^{\mu1}$. Using \bref{poinctrans} and
\bref{poincparam} these points can be described in the original
$(x^\mu)$ coordinates as follows
$$
x^\mu=\rho\ \! n^\mu(t_0) + x^\mu(t_0).
$$
By dropping the subindex of $t_0$ we get precisely the coordinate
transformation \bref{bosechange}.  This method can thus provide a
systematic way to derive the appropriate change of variables in more
complex situations, where it is less simple to guess it. In fact, we
will use a completely analogous procedure later on in order to derive
the proper generalization of \bref{bosechange} for supersymmetric
two-dimensional models with solitonic solutions.

\medskip

After the splitting between massless and massive modes one should
``integrate'' the massive modes, {\it i.e.} eliminate them by means of
their equation of motion:
\begin{equation}
\frac{\partial S}{\partial\phi}[x,\phi]=0,\quad\Rightarrow\quad
\phi=\phi[x(s)].
\label{phieq}
\end{equation}
In this way, after introducing this solution back into the original
action we will get an effective action describing the motion of
the soliton as a point particle
\begin{equation}
S_{\rm eff}[x]=S[x,\phi[x]].
\label{effact}
\end{equation}
Furthermore, any solution $x^*(s)$ of \bref{effact} will also provide
a solution of the original equations of motion through the assignment
$\phi^*= \phi[x^*]$.

The explicit realization of this programme requires some care in order
to avoid overcounting of degrees of freedom (see \cite{JJ} for a
detailed analysis of this problem). Basically, one should make sure
when deriving the equation of motion for $\phi$ that the appropriate
boundary conditions are satisfied. For example, if we assume the
worldline $x^\mu(t)$ to be the locus of zero-field spacetime points,
{\it i.e.} those satisfying
\begin{equation}
\phi(t,0)=\varphi(x^\mu(t))=0,
\label{core}
\end{equation}
(the so-called {\it core} of the field), we should take care of using
the variational principle subject to the constraint
$\delta\phi|_{\rho=0}=0$, which can be enforced by adding to the
action a Lagrange multiplier times the condition \bref{core}.  This
term, however, leads to delta-type contributions to the equation of
motion, giving rise to a non-analytic behavior for $\phi$ at the
location of the core \cite{CarGreg}.  Other definitions for the
worldline representing the soliton, {\it i.e.\ }other constraints
different than \bref{core}, are also possible and, in fact,
preferable.  The simplest amongst them is given by
\begin{equation}
\chi[\phi]=\int d\rho\;\varphi'_S(\rho)(\phi(t,\rho)-\varphi_S(\rho))=0,
\label{nocore}
\end{equation}
which ensures that $\delta\phi=\phi(t,\rho)-\varphi_S(\rho)$ is
orthogonal to the translational zero mode $\varphi'_S(\rho)$. Thus, it
guarantees that the dynamics of this zero mode is no longer described
by $\phi(t,\rho)$ but by the worldline variables $x^\mu(t)$.  This
condition can be regarded as a ``smoother'' version of the {\it core}
condition \bref{core}, in the sense that it avoids the presence of
non-analyticities at $\rho=0$.  Yet, both conditions \bref{core} and
\bref{nocore} can be seen to coalesce for large values of the soliton
mass.  So, in summary, the action used to derive $S_{\rm eff}[x]$ can
be written, using the proper time parametrization, as
\begin{equation}
S[x,\phi,g]=\int
dsd\rho\;\Delta\left[\frac1{2\Delta^2}(\partial_s\phi)^2 -\frac12
(\partial_\rho\phi)^2-\frac12(U'(\phi))^2\right]+\int ds\;
g(s)\chi[\phi],
\label{bosenewact}
\end{equation}
where $\Delta=1+\rho k$ is the determinant of the coordinate change
\bref{bosechange} and $k$ is the (signed) curvature of the worldline.

By writing the equation of motion for $\phi$ derived from
\bref{bosenewact} it is simple to show that an exact solution is given
by
\begin{equation}
\phi(s,\rho)=\varphi_S(\rho),
\label{sollike}
\end{equation}
where $\varphi_S$ stands for the static soliton solution centered at
the origin. Because of the $\rho$ dependence, this solution needs not
represent in principle a static configuration, but one that looks like
a static configuration in the co-moving frame of $x^\mu(s)$.  However,
plugging it into \bref{bosenewact} we find an effective action
describing just a free particle motion
$$
S_{\rm eff}[x]=-M_S\int ds,
$$
with a mass given by the soliton mass 
\begin{equation}
M_S=\int d\rho\;(\varphi_S'(\rho))^2.
\label{solmass}
\end{equation}

The effect of fluctuations around the soliton-like configurations
described by \bref{sollike} can be studied with the help of the
expansion
$$
\phi(t,\rho)=\varphi_S(\rho)+\zeta(t,\rho),
$$
by solving for the small fluctuations $\zeta(t,\rho)$ by means of
their own equations of motion.  This will generally produce new
contributions to the effective action, making it depart from the
trivial free propagation.

\medskip

As a summary of the procedure described in this section; in order to
explicitly derive the effective action for topological defects we have
to go through the following steps: ({\it i\,})\,perform a change of
variables in the base spacetime \bref{bosechange} so as to get a
covariant splitting between broken and unbroken directions, ({\it
ii\,})\,introduce an explicit parametrization of the topological
defect as an embedding in spacetime, $x^\mu(t)$, covariantly
describing the zero modes of the theory, and eliminate from the field
$\phi$ the dependency on these zero modes by constraining it
\bref{nocore}, and ({\it iii\,})\,solve the $\phi$ field equation
\bref{phieq} and, plugging the solution into the action, get the
effective action for the zero modes.

\section{Solitons in two-dimensional supersymmetric models}

In this section and the following one we shall generalize the above
analysis to two-dimensional supersymmetric models.  First, we will
provide an appropriate parametrization of the static (super)
domain-wall solutions. Then, after studying the relevant geometric
properties of this kind of superembeddings we will go on to discuss
the explicit derivation of the effective action.

The generic form of a supersymmetric scalar field theory in two
dimensions%
\footnote{Our conventions for the metric and gamma matrices 
are: $(g_{\mu\nu})={\rm diag}(+\;-)$ and
$\gamma^0=\sigma_2$, $\gamma^1=i\sigma_1$, $\gamma_5=\sigma_3$.}
 \cite{dVF}  
\begin{equation}
  S=\int d^2 x\;
 {1\over2}\left[(\partial_\mu\varphi)^2
  -(U'(\varphi))^2+\bar\psi i\slashed\partial\psi
   -U''(\varphi)\bar\psi\psi \right],
\label{susyaction}
\end{equation}
coupling a Majorana fermion $\psi$ and a boson $\varphi$ by means of
an arbitrary function $U(\varphi)$, exhibits supersymmetric soliton
configurations whenever the associated bosonic field theory itself
gives rise to soliton configurations as well.  Indeed, a simple
inspection to the equations of motion arising from \bref{susyaction}
\begin{eqnarray}
\partial^2\varphi+U'U''+{1\over2}U'''\bar\psi\psi=0,&&\mbox{} 
\nonumber\\
\slashed\partial\psi+iU''\psi=0,&&\mbox{}
\label{susyeqsmot}
\end{eqnarray}
shows that the static soliton and antisoliton solutions of the purely
bosonic theory,
\begin{equation}
\varphi'_{S,A}=\pm U'(\varphi_{S,A}),\quad\quad\quad\psi=0,
\label{bogo}
\end{equation}
provide also lowest-energy solutions of the supersymmetric model.
New lowest-energy solutions can now be obtained by simply applying a
SUSY transformation to the above configuration
\begin{eqnarray}
&&\delta\varphi=\bar\alpha\psi=0, 
\nonumber\\
&&\delta\psi=-i[\slashed\d\varphi_{S,A}-iU'(\varphi_{S,A})]\alpha=-2U'
(\varphi_{S,A})\alpha_\pm,
\label{susytransf}
\end{eqnarray}
where $\alpha_\pm=\frac12(1\pm i\gamma^1)\alpha$.

By looking at these transformations it is clear that only half of the
supersymmetries are actually effective when acting on a given
solution.  This implies that the initial bosonic solution ---and, in
fact, the new solutions as well--- will still be invariant under (the
other) half of the supersymmetries \bref{susytransf}. In particular,
splitting them into the $\delta_\pm$ parts, generated by $\alpha_\pm$
respectively, the soliton (antisoliton) configurations are seen to
maintain the $-$ ($+$) part of \bref{susytransf}.
Then, by applying a broken SUSY transformation onto
the soliton configuration, we will get a new
lowest-energy solution.
In other words, the flat directions of the superpotential are
parametrized by the coordinate of the soliton center of mass, $a$, and
its SUSY counterpart, $\alpha_+$, which together describe, albeit in a
non-covariant way, the location of the supersoliton in superspace.%
\footnote{We will hereafter concentrate, unless otherwise stated, on
the soliton background. Of course, everything goes through in the same
way for the antisoliton solution, after keeping track of a few
changes of sign.}  

All these features are in fact most conveniently described in
superspace. Indeed, introducing the scalar superfield
$$
  \Phi(x,\theta)=\varphi(x)+\bar\theta\psi(x)
  +{1\over2}\bar\theta\theta F(x),
$$
and the standard covariant derivative
$$
{\cal
D}_\alpha={\partial\over\partial\bar\theta^\alpha}
-i(\gamma^\mu\theta)_\alpha\partial_\mu,
$$
one can make up the manifestly SUSY-invariant action
\begin{equation}
   S=-i \int d^2xd^2\theta\left[{1\over4}\;
   \overline{{\cal D}\Phi}{\cal D}\Phi +U(\Phi)\right],
\label{superaction}
\end{equation}
which boils down to \bref{susyaction} after integration over $\theta$
and elimination of the auxiliary field $F$.  The above lowest-energy
SUSY soliton solutions can now be written in a compact form as
\begin{equation}
\Phi_S(x^\mu,\theta)=\varphi_S\left(x-a-\bar\theta_-(\theta_+-\alpha_+)
\right).
\label{sfsolution}
\end{equation}
Such a form indicates that by setting $a=0$ and $\alpha_+=0$ it will be
describing a supersymmetric soliton sitting at $x=0$ and
$\theta_+=0$. On the other hand, it is direct to show that the
coordinate combination $(x-a-\bar\theta_-(\theta_+-\alpha_+))$ present
in \bref{sfsolution} is invariant under both $t$ and $\theta_-$
translations ($\delta_-$ SUSY transformations), meaning that the
tangents to the $t$ and $\theta_-$ superspace coordinates remain as
the unbroken symmetry directions.

It is thus natural to parametrize these solutions by means of an
embedding $(X^\mu(t,\tau),\Theta(t,\tau))$, representing the location
of the domain wall in superspace. In the case of $a=\alpha_+=0$ this
embedding can be described in the simple form
\begin{equation}
X^\mu(t,\tau)=(t-t_0)\;\delta^{\mu0},\quad\quad
\Theta(t,\tau)=(\tau-\tau_0)\;\eta_-,
\label{flatsemb}
\end{equation}
where $\eta_-$ is a constant chiral bosonic spinor.

One might reasonably question why it is necessary to describe the
topological defect with a superembedding of the form $(X^\mu(t,\tau),
\Theta(t,\tau))$, since it is quite clear that representing it as
$(x^\mu(t),\theta(t))$ would already suffice for a low-energy
covariant description of the system.  The benefit of the former
description over a standard worldline parametrization lies in the fact
that the geometric picture of the derivation of the effective action
is much more transparent in terms of the super-worldline coordinates
$(t,\tau)$.  This is because, in this way, we will be able to mimic
very closely the splitting process between broken and unbroken
directions that we have used for the bosonic model. Moreover, the
standard constraints that are usually imposed for a geometrical
derivation of $\kappa$--symmetry will appear very naturally in this
setting.

\section{Supersymmetric covariant effective action}

\subsection{Geometry of domain wall superembeddings}

The presence of a second (fermionic) direction of unbroken symmetry
means that the soliton can be viewed as an extended object in
superspace, spanning along a fermionic direction.  Hence, an effective
description for it can naturally be made in terms of an embedding
describing the location of the defect in superspace:
\begin{equation}
X^\mu=X^\mu(t,\tau),\quad\quad \Theta=\Theta(t,\tau),
\label{superembedding}
\end{equation}
where the super-worldline coordinates $(t,\tau)$ parametrize the
two unbroken symmetry directions.

The remaining task should then be trying to find the appropriate
change from the field-theoretical degrees of freedom of the superfield
$\Phi$ to the superembedding $(X^\mu(t,\tau),\Theta(t,\tau))$ plus the
other (massive) modes.  In complete analogy with the bosonic case, we
should first perform a change of coordinates in superspace, {\it i.e.}
to find a covariant splitting between ``instantaneous'' broken and
unbroken superspace directions. The unbroken directions are given just
by the tangents to $t$ and $\tau$ coordinates whereas the broken ones
will be a suitable generalization of the normal vector, parametrized
by $\rho$, and a further fermionic coordinate, related to the broken
supersymmetry direction.

To proceed we will consider a super-Poincar{\'e} transformation
$(x^\mu,\theta)\rightarrow(z^\mu,\xi)$ parametrized as follows
\begin{eqnarray}
\nonumber
x^\mu&=&\Lambda^\mu_{\;\nu}(z^\nu-i\bar\xi\gamma^\nu\alpha)+a^\mu,
\\
\theta&=&S(\Lambda)(\xi+\alpha),
\label{suppoinc}
\end{eqnarray}
where $S^{-1}(\Lambda)\gamma^\mu
S(\Lambda)=\Lambda^\mu_{\;\nu}\gamma^\nu$.  We want to choose it in
such a way that the embedding $(X^\mu(t,\tau),\Theta(t,\tau))$, when
written in the new coordinates $(z^\mu,\xi)$, be (locally around a
chosen point $(t_0,\tau_0)$) represented as a flat superembedding of
the form \bref{flatsemb} obtained in the previous section. In other
words, such super-Poincar{\'e} transformation should implement the change
to a super-comoving frame for the point $(t_0,\tau_0)$ of the
superembedding.  Then, the supersymmetric analogous of the change
\bref{bosechange} will come out as relating a superspace point of
coordinates $z^\mu=\rho\delta^{\mu1}$, $\xi=\eta_+$, at the
``instant'' $(t_0,\tau_0)$ to its $(x^\mu,\theta)$ coordinates.

In order to get the explicit form of the transformation
\bref{suppoinc} we can expand $(Z^\mu(t,\tau),\Xi(t,\tau))$ ---the
expression of the superembedding in the $(z^\mu,\xi)$ coordinates---
to first order around $(t_0,\tau_0)$
\begin{eqnarray}
\nonumber
Z^\mu(t,\tau)&=&Z^\mu(t_0,\tau_0)+\dot Z^\mu(t_0,\tau_0)(t-t_0)+
Z'^\mu(t_0,\tau_0)(\tau-\tau_0)+\ldots,
\\\nonumber
\Xi(t,\tau)&=&\Xi(t_0,\tau_0)+\dot \Xi(t_0,\tau_0)(t-t_0)+
\Xi'(t_0,\tau_0)(\tau-\tau_0)+\ldots.
\end{eqnarray}
Since we want to match \bref{flatsemb} we should require
$Z^\mu(t_0,\tau_0)$ and $\Xi(t_0,\tau_0)$ to vanish.  This condition
fixes the translations and supersymmetry part of the super-Poincar{\'e}
transformation \bref{suppoinc} to be
\begin{equation}
a^\mu=X^\mu(t_0,\tau_0),\quad\quad
\alpha=S^{-1}(\Lambda)\Theta(t_0,\tau_0).
\label{suppoincprm1}
\end{equation}
Then, a further Lorentz transformation will also get $\dot
z^\mu(t_0,\tau_0)$ to point in the time direction.  This is achieved
by choosing
\begin{equation}
(\Lambda^\mu_{\;\nu})=\left(V^\mu(t_0,\tau_0)\quad N^\mu(t_0,\tau_0)
\right),
\label{suppoincprm2}
\end{equation}
where the vector $V^\mu(t,\tau)$ is defined as the unitarization of the
SUSY-invariant tangent $W^\mu(t,\tau)$
$$
V^\mu=\frac{W^\mu}{\sqrt{W^2}},\quad\quad
W^\mu=\dot X^\mu - i\bar\Theta\gamma^\mu\dot\Theta,
$$
and $N^\mu(t,\tau)$ is the unit normal, given by
$N^\mu=\epsilon^{\mu\nu} V_\nu$.

We have already fixed completely the super-Poincar{\'e} transformation,
but we are still far from having the superembedding to look like
\bref{flatsemb} at first order around $(t_0,\tau_0)$.  In addition to
super-Poincar{\'e}, we can also resort to an arbitrary
super-reparametrization of the superembedding. However, a simple
counting sufices to convince oneself that this is not enough freedom
in order to bring, even to first order, an arbitrary superembedding to
the desired form \bref{flatsemb}.  Since we are interested in
deformations of the flat domain wall solution, our focuss should be on
superembeddings that locally resemble a flat solution. This is an
indication that superembeddings describing super-domain walls cannot
be completely general ones but will be restricted by some condition.
Actually, for the sake of derivation of the change of variables, we do
not need the explicit form of this condition but only the expression
of the super-Poincar{\'e} transformation above. So we will just go on by
assuming that the superembedding $(X^\mu(t,\tau),\Theta(t,\tau)$ does
obey the required condition. Once we analyze the properties of the
change of variables it will be quite straightforward to find this
condition explicitly.

Following exactly the same steps as for the derivation of the bosonic
change of variables \bref{bosechange}, we can get the appropriate form
of the superspace change of coordinates from the expression of the
superspace points of the form $(z^\mu=\rho\delta^{\mu1}, \xi=\eta_+)$
(those spanning the instantaneous broken symmetry directions), in
terms of the original $(x^\mu,\theta)$ coordinates.  Indeed, if we
explicitate the transformation \bref{suppoinc}, with
\bref{suppoincprm1} and \bref{suppoincprm2}, we get for these points
\begin{eqnarray}
\nonumber
x^\mu &=& X^\mu(t,\tau)+\rho N^\mu(t,\tau)+i\bar\Theta(t,\tau)
\gamma^\mu\epsilon_+,
\\
\theta &=& \Theta(t,\tau)+\epsilon_+,
\label{susychange}
\end{eqnarray}
where we have already dropped the subindex `0' to identify the point
$(t,\tau)$ of the superembedding.  Here $\epsilon_+\equiv
S(\Lambda)\eta_+$ parametrizes the broken supersymmetry direction and,
because of the relation $(1-i\gamma^1)\eta_+=0$, it satisfies the
constraint
$$
\left(1+i\;\Nslash(t,\tau)\right)\epsilon_+=0.
$$
Equation \bref{susychange} is thus a change from the superspace
coordinates $(x^\mu,\theta)$ to a new set of (curved) coordinates
$(t,\tau,\rho,\epsilon_+)$.

Let us describe a few properties of this change of coordinates. First
of all, it is simple to check that both $\rho$ and $\epsilon_+$ are
invariant under supersymmetry transformations. We can also study how
\bref{susychange} is affected by a superworldline reparametrization
$(t,\tau)\rightarrow(\tilde t,\tilde\tau)$. In other words, given a
fixed superspace point $(x^\mu,\theta)$, we want to study the relation
between the $(t,\tau, \rho,\epsilon_+)$ coordinates obtained from
\bref{susychange} with the superembedding
$(X^\mu(t,\tau),\Theta(t,\tau))$ and the coordinates $(\tilde
t,\tilde\tau,\tilde\rho,\tilde\epsilon_+)$ obtained after a
superworldline reparametrization ({\it i.e.}~with $\tilde X^\mu(\tilde
t, \tilde\tau)=X^\mu(t,\tau)$ and $\tilde\Theta(\tilde t,\tilde\tau)=
\Theta(t,\tau)$).

A quick inspection shows that we must have
$\tilde\epsilon_+= \epsilon_+$, $\tilde\rho=\rho$ and that $N^\mu$,
and hence $V^\mu$ as well, must be scalar
\begin{equation}
\tilde V^\mu(\tilde t,\tilde\tau)= V^\mu(t,\tau),\quad\quad
\tilde N^\mu(\tilde t,\tilde\tau)= N^\mu(t,\tau).
\label{VNscalar}
\end{equation}
This is quite reasonable, since these quantities should depend only on
the extrinsic geometry of the superembedding and not on the way it is
parametrized.  Yet, this simple observation will suffice to determine
the condition that we alluded to above.

With this idea in mind, let us first check the transformation of
various objects under super-reparametrizations.  Consider the
SUSY-invariant generalization of the fermionic tangent vector,
$\partial_\tau X(t,\tau)$. It is given by
$$
U^\mu=DX^\mu+i\bar\Theta\gamma^\mu D\Theta,
$$
where $D=\partial_\tau-i\tau\partial_t$.  A direct computation shows
that $W^\mu$ and $U^\mu$ transform under super-reparametrizations
acording to
\begin{eqnarray}
\nonumber
W^\mu(t,\tau)&=&(\partial_t\tilde t-i\tilde\tau\partial_t\tilde\tau)
\;\tilde W^\mu(\tilde t,\tilde\tau)+\partial_t\tilde\tau\;\tilde U^\mu
(\tilde t,\tilde\tau),
\\
U^\mu(t,\tau)&=& (D\tilde t+i\tilde\tau D\tilde\tau)\;
\tilde W^\mu(\tilde t,\tilde\tau)+
D\tilde\tau\; \tilde U^\mu(\tilde t,\tilde\tau).
\label{sreptransf}
\end{eqnarray}
Since $W^\mu$ mixes in general with $U^\mu$ this implies that the
$V^\mu$ and $N^\mu$ vectors of an arbitrary superembedding
\bref{superembedding} will not be scalar under general
super-reparametrizations. By writing $U^\mu$ in the $2d$ vector basis
formed by $(V^\mu,N^\mu)$ it is clear from \bref{sreptransf} that only
when $U^\mu\propto W^\mu$ will $V^\mu$ and $N^\mu$ be scalar.
We get in this way the following (super-reparametrization
invariant) condition for the superembedding
\begin{equation}
U\cdot N=0.
\label{dwsemb}
\end{equation}
This condition is already obeyed by the flat domain wall
superembedding \bref{flatsemb}, and it is in fact the condition we have
discussed about when deriving the change of variables. Indeed,
one can explicitly check that arbitrary superembeddings satisfying
\bref{dwsemb} look, locally around any point $(t_0,\tau_0)$, like the
flat domain wall superembedding \bref{flatsemb} ({\it i.e.}, it is
possible to bring the superembedding to the form \bref{flatsemb}, at
first order around $(t_0,\tau_0)$, with a suitable choice of
super-Poincar{\'e} and super-reparametrization transformations). This is
why we will call them {\it domain wall superembeddings} and we will
hereafter assume this condition to be satisfied.

\medskip

At this point, it is possible to check that a domain wall
superembedding contains precisely the degrees of freedom that one
expects, {\it i.e.} those describing a (super)particle moving in
superspace. To show it, we can first reduce the field content of the
superembedding by restricting the general super-reparametrization
invariance with a (partial) gauge-fixing $U\cdot V=0$ which, together
with \bref{dwsemb}, implies that
\begin{equation}
U^\mu=DX^\mu+i\bar\Theta\gamma^\mu D\Theta=0.
\label{sconfc}
\end{equation}
It is clear from \bref{sreptransf} that this condition is preserved by
those restricted superdiffeomorphisms $(t,\tau)\rightarrow(\tilde t,
\tilde\tau)$
satisfying 
\begin{equation}
D\tilde t+i\tilde\tau D\tilde\tau=0,
\label{sconfeq}
\end{equation}
which can be regarded as a one-dimensional analogue of $2d$-superconformal 
transformations.
Using \bref{sconfc} we can show that the whole superembedding
$(X^\mu(t,\tau),\Theta(t,\tau))$ is completely determined in terms of
the worldline supercoordinates $(x^\mu(t),\theta(t))$.
Indeed, expanding the embedding superfields as
\begin{eqnarray}
\nonumber
X^\mu(t,\tau)&=&x^\mu(t)+i\tau\psi^\mu(t),
\\
\Theta(t,\tau)&=&\theta(t)+\tau\lambda(t),
\label{decomposition}
\end{eqnarray}
and imposing the constraint \bref{sconfc} we are led to the
following conditions
\begin{eqnarray}
\nonumber
\psi^\mu&=&-\bar\lambda\gamma^\mu\theta,
\\
\bar\lambda\gamma^\mu\lambda&=&\dot 
x^\mu-i\bar\theta\gamma^\mu\dot\theta.
\label{components}
\end{eqnarray}
Multiplying \bref{components} with $n^\mu(t)$, the unit vector
orthogonal to $w^\mu(t)\equiv\dot x^\mu-i\bar\theta\gamma^\mu\dot\theta$,
we find that $\lambda$ has to be chiral with respect to the splitting
$\lambda_\pm=1/2\;(1\mp i\;\slashed n)\lambda$.
We get the correct chirality for $\lambda$ by noting that
\bref{decomposition} should reduce to \bref{flatsemb} in the limit of
a static soliton, which implies that $\lambda=\lambda_-$.

To complete the analysis of the low energy degrees of freedom we
should prove that the gauge freedom generated by the restricted
superdiffeomorphisms $(t,\tau)\rightarrow(\tilde t,\tilde\tau)$ 
satisfying \bref{sconfeq} corresponds to just worldline
reparametrizations plus $\kappa$--symmetry. Expanding the infinitesimal
transformations as
\begin{eqnarray}
\nonumber
\delta t&=&a(t)-i\tau\beta(t),
\\\nonumber
\delta\tau&=&\alpha(t)+\tau b(t),
\end{eqnarray}
we see that condition \bref{sconfeq} implies $\beta=-\alpha$ and
$b=-\dot a/2$.  Being $X^\mu(t,\tau)$ and $\Theta(t,\tau)$ scalar
superfields, these restricted superdiffeomorphisms induce the following
transformations on the component fields $x^\mu(t)$ and $\theta(t)$
\begin{eqnarray}
\nonumber
\delta x^\mu&=&a\;\dot x^\mu+i\bar\theta\gamma^\mu\alpha\lambda,
\\\nonumber
\delta\theta&=&a\;\dot\theta+\alpha\lambda.
\end{eqnarray}
We thus identify $a$ and $\kappa\equiv\alpha\lambda$ respectively as
the generators of infinitesimal worldline repa\-ram\-etriza\-tions and
$\kappa$--symmetry transformations.

This sort of geometrical interpretation for $\kappa$--symmetry has
been previously observed by various authors
\cite{STV,Gauntlett,IvaKap2} and further pursued in \cite{BSTPV}.  The
starting point in these papers is to consider a superembedding
satisfying some sort of geometrical constraint ---the so-called {\it
geometrodynamical constraint}--- from which one finds, as we have
reproduced above, that invariance under those restricted
superdiffeomorphisms satisfying the constraint is equivalent to
worldline reparametrizations plus $\kappa$--symmetry invariance.  We
have seen that our analysis on the effective description of the
soliton dynamics naturally accomodates within this formulation. Here
we regard the superembedding $(X^\mu(t,\tau),\Theta(t,\tau))$ as the
locus of a topological defect and give a neat interpretation of the
constraint \bref{sconfc} as the result of, first, selecting a definite
type of superembeddings ---the domain wall superembeddings, which
locally in an appropriate frame resemble a static solution--- and,
second, reducing invariance under general super-reparametrizations to
the restricted superdiffeomorphisms satisfying equation
\bref{sconfeq}.  In this sense, we have traced here the origin of the
geometrodynamical constraints leading to $\kappa$--symmetry by
studying the interplay between the effective model and the underlying
field theory.

\medskip

Before we go on to explicitly obtain the effective action it will be
useful to derive a few geometric relations obeyed by domain wall
superembeddings in the gauge $U\cdot V=0$.  Taking the covariant
derivative of equation \bref{sconfc}, $U^\mu(t,\tau)=0$, we
immediately get
\begin{equation}
V^\mu=\nabla\bar\Theta\gamma^\mu\nabla\Theta,
\label{vexpr}
\end{equation}
where we have introduced the scalar covariant derivative
$$
\nabla=\frac 1{E^{1/2}} D,
$$
with $E=\sqrt{W^2}$. Now, multiplying \bref{vexpr} by $N^\mu$ and
using the orthogonality condition $N\cdot V=0$ we find that $\nabla
\Theta$ must be a chiral spinor. Since it has to tend, in addition, to
the static solution \bref{flatsemb} in the flat superembedding limit,
we conclude that it has to satisfy the chirality constraint
\begin{equation}
(1-i\;\Nslash)D\Theta=0.
\label{chirality}
\end{equation}

It is now simple to derive the following generalization of Frenet
equations:
\begin{eqnarray}
\nonumber
\nabla V(t,\tau)&=&{\cal K}\;N(t,\tau),
\\\nonumber
\nabla N(t,\tau)&=&{\cal K}\;V(t,\tau),
\end{eqnarray}
where ${\cal K}(t,\tau)$ plays the role of a (fermionic) curvature
characterizing the superembedding.  Owing to the relations
\bref{sconfc} and \bref{chirality} we can express all these objects in
terms of $\Theta(t,\tau)$ alone:
\begin{eqnarray}
\nonumber
E&=&-iD\bar\Theta\gamma_5D\Theta,
\\\nonumber
{\cal K}&=&2i\nabla^2\bar\Theta\nabla\Theta,
\\\nonumber
V^\mu&=&\nabla\bar\Theta\gamma^\mu\nabla\Theta,
\\\nonumber
N^\mu&=&\nabla\bar\Theta\gamma^\mu\gamma_5\nabla\Theta.
\end{eqnarray}

{}From equation \bref{chirality} we notice that $D\Theta$ and
$\epsilon_+$, the spinor parameter appearing in the superspace change
of variables \bref{susychange}, have opposite chiralities. Since
$\epsilon_+$ describes only one degree of freedom, it is possible to
rewrite it with the help of $\nabla\Theta$ as
\begin{equation}
\epsilon_+=\gamma_5\nabla\Theta\;\sigma,
\label{sigmadef}
\end{equation}
where $\sigma$ is a scalar SUSY-invariant fermionic parameter.

Taking all of this into account, the final form of the change of
variables in superspace is
\begin{eqnarray}
\nonumber
x^\mu &=& X^\mu(t,\tau)+\rho N^\mu(t,\tau)+i\bar\Theta(t,\tau)
\gamma^\mu\gamma_5\nabla\Theta(t,\tau)\;\sigma,
\\
\theta &=& \Theta(t,\tau)+\gamma_5\nabla\Theta(t,\tau)\;\sigma,
\label{susychange2}
\end{eqnarray}
where the new superspace parametrization is made in terms of 
$(t,\tau,\rho,\sigma)$.

\subsection{Derivation of the effective action}

In order to find the effective action we will first expand the
superfield action \bref{superaction} around a soliton-like
configuration and then implement the above change of variables
\bref{susychange2}. We thus consider the following expansion for the
superfield $\Phi(x,\theta)$
$$
\Phi(x,\theta)=\varphi_S(\rho)+\zeta(t,\tau,\rho,\sigma),
$$
where $\zeta(t,\tau,\rho,\sigma)$ will be taken as a small
perturbation.  Just as in the purely bosonic model, the soliton-like
configuration $\varphi_S(\rho)$ need not generally represent a true
static soliton configuration but one that looks static in a frame
traveling with the defect.  Notice that, being the variable $\rho$
inert under supersymmetry, this is a SUSY-invariant decomposition. It
is also simple to show that $\rho$ tends to the coordinate combination
$(x-\bar\theta_-\theta_+)$ as $(X^\mu(t,\tau),\Theta(t,\tau))$
approaches the flat superembedding \bref{flatsemb}, implying that
$\varphi_S(\rho)$ tends to the flat solution \bref{sfsolution}
with $a=\alpha_+=0$.

We thus have a splitting of the action in powers of the perturbation
superfield $\zeta$
\begin{equation}
S=S_0+S_1+S_2+\ldots,
\label{Sexpansion}
\end{equation}
where
\begin{eqnarray}
\nonumber
S_0&=&-i\int d^2xd^2\theta\;\left(\frac14\overline{{\cal D}\varphi_S}{\cal D}
\varphi_S+U(\varphi_S)\right),
\\\nonumber
S_1&=&-i\int d^2xd^2\theta\;\left(\frac12\overline{{\cal D}\varphi_S}{\cal
D}\zeta+U'(\varphi_S)\zeta\right),
\\
S_2&=&-i\int d^2xd^2\theta\;\left(\frac14\overline{{\cal D}\zeta}{\cal D}
\zeta+\frac12U''(\varphi_S)\zeta^2\right).
\label{actexp}
\end{eqnarray}

We should now perform the change of variables \bref{susychange2} in
these integrals.  There is, however, an important point that has to be
taken into account when making a change of variables involving
fermionic coordinates and which is relevant to our derivation of the
effective action.  It refers to the fact that the computation of a
superspace integral can produce a wrong answer if, prior to
integration, one performs a change of variables involving a generic
mixing of bosonic and fermionic degrees of freedom.  This is a known
problem and it was already encountered by Ivanov and Kapustnikov
\cite{IvaKap1} in their search for a non-covariant gauge-fixed version
of the effective action.  We can illustrate this point with the
following simple example.  Consider an integral of the form $\int
dxd\theta_1d\theta_2 F(x-\theta_1\theta_2)$, where $x$ is a single
real variable and $\theta_1$, $\theta_2$ are a pair of real Grassmann
variables. If we perform the integration directly we get
$$
\int dxd\theta_1d\theta_2 F(x-\theta_1\theta_2)=\int
dx\;F'(x)=F(\infty)-F(-\infty).
$$
However, if we would make a change of variable $x\rightarrow \rho=
x-\theta_1\theta_2$, with a unit Jacobian, before the integration, we
would get
$$
\int d\rho d\theta_1d\theta_2 F(\rho)=0,
$$
in contradiction with the correct result whenever $F(\infty)\neq
F(-\infty)$.

This sort of problems arise in integrals involving functions which do
not fall to zero at infinity and the discrepancy between both
calculations is always a boundary term. This is precisely the
situation with our soliton configurations $\varphi_S$, which tend to
non-zero vacuum values at space infinity. Because of this, changing to
the new variables $(t,\tau,\rho,\sigma)$ and integrating over $\rho$
and $\sigma$ to get the effective action at the lowest
($\zeta$-independent) order, $S_0$, would not produce the right answer
for this term. On the other hand, it is clear that this problem will
not appear for the other ($\zeta$-dependent) contributions, $S_1$,
$S_2,\ldots$, because the $\zeta$-field boundary conditions force
these integrals to tend to zero at large distances, thus preventing
the presence of any boundary-related terms.

The way to solve this problem is to re-express, in the $S_0$ part,
the coordinate $\rho$ in terms of a new coordinate $\tilde\rho$
which tends to the pure space coordinate, $x$, as the superembedding
approaches the flat configuration \bref{flatsemb}.
The expression of $\tilde\rho$ can be obtained by rewritting the
bosonic part of the change of variables \bref{susychange2} as
$$
x^\mu=X^\mu(t,\tau)+\beta V^\mu(t,\tau)+\tilde\rho N^\mu(t,\tau),
$$
which implies the relation
$$
\tilde\rho=\rho+\bar\Theta\gamma_5\nabla\Theta\;\sigma.
$$
Expressing the $S_0$ part of the action \bref{actexp} in terms of
$\tilde\rho$ we immediately see that the whole contribution comes from
the potential term. Explicitly,
$$
S_0=i\int dtd\tau d\tilde\rho
d\sigma\;E^{1/2}\left[1+\sigma{\cal K}+i(\tilde\rho-\bar\Theta
\gamma_5\nabla\Theta\sigma)\nabla{\cal K}\right]
\;U\left(\varphi_S(\tilde\rho-\bar\Theta
\gamma_5\nabla\Theta\sigma)\right),
$$
where we have used the expression of the super-Jacobian for the change
of variables \bref{susychange2} which is found to be
$$
J=-E^{1/2}(1+\sigma{\cal K}+i\rho\nabla{\cal K}).
$$

After integrating over $\sigma$ and $\tilde\rho$ we find that all
curvature-dependent terms pack up in a total time derivative, so that
they do not contribute at this order.  We get in this way the correct
expression for the lowest-order contribution to the effective
action. It is given by
\begin{equation}
S_0=M_S\int dtd\tau\;E^{1/2}\;\nabla\bar\Theta\gamma_5\Theta,
\label{S0act}
\end{equation}
where $M_S$ is the soliton mass defined in \bref{solmass}.
This action is manifestly invariant under supersymmetry and under
restricted super-diffeo\-mor\-phisms.  It has been previously obtained
by Gauntlett \cite{Gauntlett} and by Ivanov and Kapustnikov \cite
{IvaKap2} as the simplest model that can be constructed in this
geometrical setting for $\kappa$--symmetry.

One can resort to its component form in order to identify the nature
of this action.  Plugging the superfield expansions
\bref{decomposition} and the constraints \bref{components} into $S_0$
we can write everything in terms of $x^\mu(t)$ and $\theta(t)$. After
integration over $\tau$ we find
\begin{equation}
S_0[x,\theta]=-M_S\int dt\left(\sqrt{(\dot
x-i\bar\theta\gamma\dot\theta)^2}+\bar\theta\gamma_5\dot\theta\right),
\label{spartact}
\end{equation}
which is nothing but the free massive superparticle action, invariant
under $\kappa$--symmetry by virtue of the Wess-Zumino term
$\bar\theta\gamma_5\dot\theta$.

\medskip

The terms $S_1$ and $S_2$ in the expansion of the action
\bref{Sexpansion}, which are respectively linear and quadratic in
$\zeta$, can be found after some tedious but straightforward
algebra. After integration over $\sigma$ we get the following
expression for $S_1$
$$
S_1[x,\theta,\eta]=\int dtd\tau E^{1/2}\int d\rho\;i
\varphi'_S(\rho)\eta(t,\tau,\rho),
$$
where the superfield $\eta$ comes from the $\sigma$ expansion of the
superfield $\zeta$
$$
\zeta(t,\tau,\rho,\sigma)=\eta(t,\tau,\rho)+\sigma\varepsilon
(t,\tau,\rho).
$$

The form of the linear term is giving us a hint of an appropriate
definition for the location of the domain wall associated with a given
$\Phi$ configuration.  A possible choice would be the {\it core} of
$\Phi$, satisfying $\Phi(X(t,\tau), \Theta(t,\tau))=0$ but a probably
better one is to take the associated superembedding as the one
satisfying
\begin{equation}
\int d\rho\;i\varphi'_S(\rho)\eta(t,\tau,\rho)=0.
\label{susygf}
\end{equation}
One can check that this condition provides a unique assignation of a
superembedding to a given superfield configuration,
$\Phi(x,\theta)\rightarrow(X(t,\tau), \Theta(t,\tau))$.  Moreover,
this condition eliminates the zero modes from $\eta$, since they are
already described by the superembedding.  It is also manifestly
invariant under restricted superdiffeomorphisms. Thus, it preserves
worldline and $\kappa$--symmetry invariance. It can be shown, in
addition, that this choice merges with the {\it core} definition as
the mass $M_S$ of the soliton tends to infinity.

Condition \bref{susygf} should be enforced in the effective action
before attempting to solve for the massive modes and obtain the
effective action for the zero modes. We have already discussed this
issue for the simpler bosonic case and the same reasoning applies here
as well.  A thorough discussion of this technical point can be found
in reference \cite{JJ}.  The bottom line is that we may absorb the
linear part of the action $S$, {\it i.e.}~the whole of $S_1$, in a
Lagrange multiplier term enforcing the condition \bref{susygf}.

\medskip

Taking all of this into account and including also the contribution
from $S_2$ we find the following form of the action to second order in
the perturbations
\begin{eqnarray}
\nonumber
S&=&-M_S\int dt\left(\sqrt{(\dot x-i\bar\theta\gamma\dot\theta)^2}
+\bar\theta\gamma_5\dot\theta\right)
+\int dtd\tau\;\Lambda(t,\tau)\;\int d\rho\;i\varphi'_S(\rho)
\eta(t,\tau,\rho)
\\\nonumber
&&+\int dtd\tau\;E^{1/2}\int d\rho\left[-\frac12\nabla\eta\nabla^2\eta
+(1+i\rho\nabla{\cal K})\left(-\frac12\varepsilon\nabla\varepsilon
-i\varepsilon\eta'+iU''(\varphi_S)\eta\varepsilon\right)\right.
\\\nonumber
&&\quad\quad\quad\quad\quad\quad\quad\quad\quad
\left.-{\cal K}\left(\frac12\varepsilon\nabla\eta-\frac
i2U''(\varphi_S)\eta^2-\frac{i\rho}{2(1+i\nabla{\cal K}\rho)}
(\nabla^2\eta)^2+\frac{i\rho}2\varepsilon\nabla^2\varepsilon
\right)\right],
\end{eqnarray}
where $\Lambda(t,\tau)$ is a Lagrange multiplier superfield.

Just as in the bosonic case, we can see that
$\Phi(x,\theta)=\varphi_S(\rho)$, that is $\zeta=0$, is already a
solution of the equations of motion for the massive modes, leaving out
the superparticle action \bref{spartact} as the only contribution to
the low energy effective action. Non-trivial boundary conditions,
however, are expected to generate other (non-trivial) solutions for
$\zeta$, giving rise to higher-order $\kappa$--invariant corrections to
the basic superparticle term.

\section{Conclusions and outlook}

We have extended in this paper the effective action method to the
domain of $2d$ supersymmetric scalar field theories.  In this way, we
have obtained a SUSY- and $\kappa$--invariant expansion for the low
energy action that can be used to derive higher-order corrections to
the basic superparticle action for topological defects.  Beyond the
interest of such an expansion, we would like to draw attention to other
potentials of this deductive approach.  Thus, for example, in our
method we are able to trace very closely the origin of
$\kappa$--symmetry from the underlying field theory action. We have
connected with the geometrical interpretation of this symmetry
previously proposed by several authors. The geometrodynamical
constraint, $U^\mu=0$ in our model, which is at the root of this
interpretation, has been understood in a very natural way from within
our approach.  Part of this constraint, $U\cdot N=0$, serves to select
the so-called {\it domain wall superembeddings} ---those that locally
resemble a flat domain wall. The rest of the constraint, $U\cdot V=0$,
acts as a gauge-fixing, reducing the general super-reparametrization
invariance down to a subset giving rise to the $\kappa$--tensor
calculus.

The above analysis raises our hope that a similar construction can be
used for higher-dimensional supersymmetric theories, aiming also at a
general $\kappa$--tensor calculus in arbitrary
dimensions. Unfortunately, a straightforward generalization of these
techniques to the closest higher-dimensional objects, that is, the
$4d$ Nielsen-Olesen supervortex and the domain wall of the $4d$
Wess-Zumino model ---respectively described by the Green-Schwarz
superstring and the supermembrane action--- is presently failing to
produce the expected answers, due to further subtleties not present in
the $2d$ model just analyzed.

Thus, for instance, when trying to describe the four-dimensional
supermembrane it is simple to check that the appropriate
generalization of equation \bref{sconfeq}, $D\tilde t+i\tilde\tau
D\tilde\tau=0$, does the job and provides the required $3d$
reparametrizations and $\kappa$--transformations.  However, the
geometrodynamical constraint alone ---which comes out in an analogous
way as in two dimensions--- is not enough to constraint the field
contents of the theory to the required one.  This fact seems to
indicate the existence of further geometrical constraints, yet
unknown, in order to properly characterize $4d$ domain wall
superembeddings.

In conclusion, although our method provides a fairly complete picture
of the geometry of the $2d$ domain wall superembeddings and of its use
in deriving low energy effective actions for SUSY topological defects,
it will still require further investigation before being generalized
to higher dimensions. As a byproduct of this extension, one should
expect a better understanding of the geometry of higher-dimensional
superembeddings and a general formulation of the long-standing problem
of $\kappa$--tensorial calculus.  Work in the directions just sketched
is already in progress.

\section*{Acknowledgements}

We would like to thank D. Mateos, J. Sim{\'o}n and F. Zamora for valuable
comments and suggestions.  This work has been supported in part by a
CICYT contract AEN95-0590.  J.R. also thanks Ministerio de Educaci{\'o}n y
Ciencia of Spain for financial support.

\end{document}